\documentstyle[manuscript,prd,aps]{revtex}
\begin{document}
\draft
\tighten
\title{LENSING PROPERTIES OF LIGHTLIKE CURRENT CARRYING COSMIC STRINGS}
\author{Jaume GARRIGA$^1$ and Patrick PETER$^{1,2}$}
\address{\hfill\\
$^1$Department of Applied Mathematics and Theoretical Physics,\\
University of Cambridge, Silver Street, Cambridge CB3 9EW (UK)\\
\hfill\\
$^2$D\'epartement d'Astrophysique Relativiste et de Cosmologie,\\
Observatoire de Paris-Meudon, UPR 176 CNRS, 92195 Meudon cedex (France)}
\date{\today}
\preprint{DAMTP-R94/10, gr-qc/9403025}
\maketitle
\begin{abstract}
The lensing properties of superconducting cosmic strings endowed with
a time dependent pulse of lightlike current are investigated. The
metric outside the core of the string belongs to the $pp$--wave class,
with a deficit angle. We study
the field theoretic bosonic Witten model coupled
to gravity, and we show that the full metric (both outside and
inside the core) is a Taub-Kerr-Shild generalization of that for the
static string with no current. It is shown that the double image due
to the deficit angle evolves in an unambiguous way as a pulse of
lightlike current passes between the source and the observer.
Observational consequences of this signature of the existence of
cosmic strings are briefly discussed.
\end{abstract}

\newpage
\section*{Introduction.}

Most of the gravitational properties of cosmic strings~\cite{kibble}
and superconducting cosmic strings~\cite{witten} have so far been
calculated in the special framework of zero~\cite{vilenkin,pp1} or
time independent currents~\cite{gravt0,ppuy} so that it is widely
believed that observation of strings in the universe might be achieved
through the now well known double image effect, i.e. the deflection
associated with the missing angle in the conical metric surrounding a
string. An interesting observation would therefore consist in two
exactly identical stars (or galaxies) separated by a fixed angle
(usually assumed of the order $10^{-6}$rad~$\sim 2.4''$, i.e., for
strings generated at the grand unified phase transition), or perhaps
an aligned series of such twin images~\cite{hu}. However, because the
probability that two nearly identical stars be separated in the sky by
pure coincidence in exactly the way one expects for cosmic strings is
rather high, there is in fact a demand for a more convincing
signature. As we shall see, a lightlike current may in principle
provide such a signature.

In this paper, we study the gravitational field created by a lightlike
current-carrying string. For the case of timelike and spacelike
currents, which have been studied before~\cite{gravt0,ppuy}, all known
solutions are stationary~\cite{ref:20}. In the lightlike case,
however, we shall see that the stationary solution has peculiar
asymptotic behaviour, in the sense that all light rays are
gravitationally bound by the string. Therefore, the problem of asymptotic
light deflection is not well posed in this metric. This is just as
well, since in
realistic situations, we do not expect the current to be stationary.
In particular we can consider a pulse of lightlike current travelling
along the string at the speed of light. Such pulses may be generated
when a long string interacts with a bounded external source. In this
case, the motion of the double image with time due to the
nonstationary gravitational field must be examined.

The article is organized as follows: in a first section, we derive the
metric surrounding a lighlike current carrying cosmic string by
boosting a spacelike current metric, and show how the resulting
gravitational field can be generalized to a time-dependent exact
solution of the Einstein-Maxwell equations. In section~\ref{Wit} we
address the question of finding the corresponding solutions in the
underlying field theoretic model. For definiteness, we use the bosonic
Witten model~\cite{witten} coupled to gravity. We show that the
introduction of a lightlike current does not modify the vortex
profile.  Also, the effect of the current on the metric is accounted
for by means of a single function which satisfies a linear
differential equation.  In order to understand the effect of a
lightlike current on a possible observation of light deflection by a
string, we use the fact that our solution can in fact account for
time-varying fields, so we consider a finite pulse of such a current.
We then work out analytically the characteristic features of the
geodesic motion in this metric, and eventually calculate numerically,
in the last section, the actual effective motion (meaning as seen by
a geodesic observer) of a point source initially aligned with a
string and an observer when a pulse of lightlike current passes
between the source and the observer. We conclude by considering
the astrophysical observational possibilities and discussing a
possible mechanism for generating pulses of lightlike current in
cosmic strings.

\section{The Exterior Metric.}

In this section, we shall obtain the gravitational field surrounding a
cosmic string carrying a lightlike current, treating the source as an
infinitely thin distribution. One easy way to do so is by the limiting
procedure~\cite{boostmeth} of boosting the metric of the spacelike
current case~--~which is known~\cite{gravt0,ppuy,Lwitten}~--~to the
speed of light in the direction parallel to the string (alternatively,
one could also boost the timelike current metric, both procedures
being equivalent).

In the case of spacelike current, for a static cylindrically symmetric
configuration with the string lying along the $z-$axis, the metric
outside the core of the string is~\cite{gravt0,Lwitten}
\begin{equation} ds^2 = g(-dt^2+dr^2)+r^2f\gamma^2d\theta^2+{dz^2\over
f}. \label{metricini}\end{equation}
Here, $f\equiv [c_1 (r/r_\sigma)^m +c_2 (r/r_\sigma)^{-m}]^2$ and
$g=(r/r_\sigma)^{2m^2} f$, where $r_\sigma$ is the thickness of the core
of the string. The nonvanishing component of the electromagnetic field
is given by
\begin{equation} F_{rz} = {2m\sqrt{c_1 c_2}\over G^{1/2} r}
f,\label{elecini} \end{equation} where $G$ is Newton's constant. The
various parameters $c_1$, $c_2$, $m$ and $\gamma$ should be
determined, in principle, in terms of the microphysical parameters
characterizing the vortex by matching the exterior
metric~(\ref{metricini}) with the solution of Einstein's equations
inside the core of the string (i.e., for $r<r_\sigma$). In practice,
this can be difficult if the gravitational field is strong. However,
for weak fields, one can arrive at the relations~\cite{ppuy}
\begin{eqnarray} m^2 &=& 4GI^2 +{\cal O}(G^2), \nonumber\\
c_1 &=& {1\over 2}[1+{G^{1/2}\over I}(U-T-I^2)]+{\cal O}(G),\nonumber\\
\null & & \label{match} \\
c_2 &=& {1\over 2}[1-{G^{1/2}\over I}(U-T-I^2)]+{\cal O}(G),\nonumber\\
\gamma &=& 1-4G(U+{1\over 2}I^2)+{\cal O}(G^2).\nonumber
\end{eqnarray}
Here, $U$, $T$ and $I$ are, respectively, the flat space values of the
string's energy per unit length, tension and current (see, e.g.,
Refs.~\cite{pp1,ppuy} for a discussion of the limiting procedure
consisting in taking well-defined flat space integrals of the stress
energy tensor and current as the source of the weak gravitational
field).

Treating the string as an idealized distribution of zero thickness,
the components of the string's energy momentum tensor are given by
$T^S_{\mu\nu} = \tau _{\mu\nu}\delta (x) \delta (y)$, [$\mu ,\nu =
t,x,y,z]$, where~\cite{gravt0}
\begin{equation} \tau_{\mu\nu} = \hbox{diag}\,
(U,{I^2\over 2},{I^2\over 2},-T). \label{taumunu}\end{equation}

A boost in the $z$ direction
\begin{eqnarray} z' &=& z \cosh \Lambda + t\sinh \Lambda ,\nonumber\\
\null & & \label{boost}\\
t' &=& t \cosh \Lambda + z\sinh \Lambda ,\nonumber
\end{eqnarray}
transforms Eq.~(\ref{metricini}) into
\begin{equation} ds^2 = gdr^2+r^2f\gamma^2d\theta^2+\alpha ({1\over f}
-g) (dz^2+dt^2-2\sqrt{{\beta\over\alpha}}dzdt) + gdz^2-{1\over f}dt^2,
\label{metint}\end{equation}
where $\alpha \equiv \cosh ^2 \Lambda$, $\beta \equiv \sinh ^2 \Lambda$ and
we have dropped the prime on the coordinates. Also the current, which
was purely spacelike $J^\mu = (0,0,0,I)\delta (x) \delta (y)$,
[$\mu =t,r,\theta ,z$],
transforms into $J'^\mu = (\sqrt{\beta} I, 0,0,\sqrt{\alpha}I) \delta
(x) \delta (y)$, and
becomes null as $\alpha \to \infty$. The $(t,z)$ components of the
stress tensor~(\ref{taumunu}) transform into
\begin{equation} \tau '_{ab} = \hbox{diag}\, (U,-T) + (U-T)
\left(
\matrix{\beta & -\sqrt{\alpha \beta} \cr -\sqrt{\alpha \beta} & \beta}
\right) \label{tauab}\end{equation}
$(a,b = z,t)$, aquiring non diagonal pieces. It is
clear that if we want the components of $J'^\mu$ and $\tau'_{ab}$ to
be finite, the limit $\alpha\to\infty$ can only be taken by letting
$I$ and $(T-U)$ go to zero at the same time, in such a way that the
products
\begin{equation} I'_z \equiv \sqrt{\alpha} I \ \ \ \hbox{and} \ \ \
(U'-T') \equiv (\beta -{1\over2}) (U-T) \label{finite}\end{equation}
remain finite.

Upon so doing, we find
\begin{equation} ds^2=-dudv-V(r)du^2+dr^2+r^2\gamma^2d\theta^2
\label{metfin} \end{equation}
where we have used null coordinates $u\equiv t-z$ and $v=t+z$.
Using Eqs.~(\ref{match}) and~(\ref{finite}), the function $V(r)$ can
be written as
\begin{equation} V(r) = 8G(U'-T')\ln {r\over r_\sigma} + 8GI_z'^2
\ln ^2 {r\over r_\sigma} ;\label{potential}\end{equation}
also we can write
\begin{equation} \gamma = 1-4G(U'+T').\label{gam}\end{equation}
{}From Eq.~(\ref{elecini}), the electromagnetic tensor in the new frame
is
\begin{equation} F'_{ru}=-{2I'_z\over r},\label{newF}\end{equation}
and the energy momentum of the string is given by
\begin{equation} \tau'_{\mu\nu} = {U'+T'\over 2} \hbox{diag}\, (1,0,0,-1)
+ (U'-T') \delta_{\mu u} \delta_{\nu u},\label{newtau}\end{equation}
where we have used Eqs.~(\ref{tauab}) and~(\ref{finite}).

Typically~\cite{num-witt}, the degeneracy $(U-T)$ is of order
$I^2/e^2$, where $e$ is the electromagnetic coupling. Therefore, from
Eq.~(\ref{finite})
$$ U'-T'\sim {I_z'^2\over e^2} .$$
In cosmological applications, $\ln (r/r_\sigma)\sim 100$. Since $e^2
\sim 10^{-2}$, both terms in Eq.(\ref{potential}) will be of the same
order of magnitude.

Although the metric~(\ref{metfin},\ref{potential}) has been obtained
using the linearized relations~(\ref{match}), we shall see that it is
actually a solution of the full Einstein-Maxwell equations, with
electromagnetic field given by Eq.~(\ref{newF}). This fact, which is
also encountered in the Aichelburgh--Sexl case~\cite{boostmeth}, has a
simple mathematical explanation. As $\alpha\to\infty$, we let $I$ and
$(T-U)$ to zero, and so we are boosting a metric for which (in the
limit $\alpha\to\infty$) the linear approximation becomes exact.

For later use, it is convenient to generalize
Eq.~(\ref{metfin}) to the case where, instead of having a stationary
null current flowing along the string, we have a localized pulse of
current which is travelling in the positive $z$ direction at the speed
of light. The metric~(\ref{metfin}) belongs to the class known as
$pp-$waves~\cite{shock}
\begin{equation} ds^2=-dudv-H(u,r)du^2+dr^2+r^2\gamma^2d\theta^2.
\label{pp}\end{equation}
For the metric~(\ref{pp}), the only nonvanishing component of the
Ricci tensor outside the core is
\begin{equation} R_{uu}= {1\over 2} \Box H(u,r) = 8\pi G T_{uu}.
\label{E} \end{equation}
The only component of the electromagnetic energy momentum tensor is
\begin{equation} T^{e.m.}_{uu} = {1\over 4\pi}F_{ur}F^{ur} =
{p^2(u)\over \pi} {I_z ^{\prime 2}\over r^2} .\end{equation}
Taking the ansatz
\begin{equation} H(u,r) = p^2 (u) V(r), \label{Hur}\end{equation}
with $V(r)$ given by Eq.~(\ref{potential}), it is clear that the
Einstein equation~(\ref{E}) is satisfied outside the core.

In the core of the string, our metric~(\ref{pp}) should be matched
with an interior solution. Instead of that, we shall treat the source
as infinitely thin, with support only at $r=0$. In a space with a
deficit angle $2\pi (1-\gamma )$, we have $$\widetilde \Delta \ln r =
2\pi\gamma \delta (x) \delta (y),$$ where $(x,y)=r(\cos\theta ,
\sin\theta )$ and $\widetilde \Delta$ is the Laplacian in the
($r,\theta$) plane. Then, one can see that the source corresponding
to~(\ref{pp}-\ref{Hur}) has the form
\begin{equation} \tau'_{\mu\nu} = {U'+T'\over 2} \hbox{diag}\,
(1,0,0,-1) + p^2(u)\gamma (U'-T')\delta_{\mu u}\delta_{\nu u}.
\end{equation}
The first term, corresponding to the Goto-Nambu string, is responsible
for the deficit angle. The second term, caused by the nondegeneracy
$(U' - T')$, is modulated by the function $p(u)$ that gives the
profile of the pulse of current. From $\nabla_\mu F^{\mu\nu} = 4\pi
J^\nu$, we have
$$J^\nu = I'_z \gamma p(u) (1,0,0,1) \delta (x) \delta (y).$$
If $p(u)$ has compact support, the metric~(\ref{pp})
represents the gravitational field of a pulse of current travelling in
the positive $z$ direction.

\section{Witten model coupled to gravity}\label{Wit}

In the previous section, we treated the superconducting string as an
infinitely thin distribution along which a pulse of lightlike current
flows. The question remains, however, of whether it is possible to
find the corresponding solutions in the underlying field theoretic
model coupled to gravity.

For definiteness, we shall consider the Witten bosonic
model~\cite{witten}, consisting of two complex scalar fields and
associated U(1) gauge fields ($\Phi ,B_\mu$) and ($\Sigma ,A_\mu$).
The Lagrangian is given by
\begin{equation} {\cal L} = -{1\over 2} |D_\mu \Phi |^2 - {1\over 2}
|D_\mu \Sigma |^2 -{1\over 16\pi}F_{\mu\nu}F^{\mu\nu}
-{1\over 16\pi}H_{\mu\nu}H^{\mu\nu} - {\cal V}(|\Phi |,|\Sigma |)
.\label{lagwit} \end{equation}
Here, $D_\mu \Phi = (\nabla _\mu +iqB_\mu)\Phi$,
$D_\mu \Sigma = (\nabla _\mu +ieA_\mu)\Sigma$, $F_{\mu\nu} =
\partial_\mu A_\nu - \partial_\nu A_\mu$ and $H_{\mu\nu} =
\partial_\mu B_\nu - \partial_\nu B_\mu$. The potential $${\cal V}
(|\Phi |,|\Sigma |) = {\lambda _\phi\over 8}(|\Phi |^2 - \eta^2)^2 +
\zeta (|\Phi |^2 -\eta^2)|\Sigma |^2 +{\lambda _\sigma\over 4} |\Sigma
|^4 + {m_\sigma ^2 \over 2} |\sigma |^2$$ is chosen so that the field
$\Phi$ undergoes spontaneous symmetry breaking. Since the vacuum
manifold is nontrivial, $\Phi$ will admit vortex (string-like)
solutions. $\Sigma$ develops a condensate in the core of the string,
but vanishes ouside. The field $A_\mu$ is then identified with
electromagnetism, which is unbroken outside the string.

To find the gravitational field of a superconducting cosmic string
with lightlike current in the model~(\ref{lagwit}), we shall proceed
in two steps. First, we shall consider the metric for the static and
cylindrically symmetric configuration in which no current is flowing
in the string. This configuration is also invariant with respect to
boosts parallel to the symmetry axis~\cite{garfinkle}. Then, we shall see
that the metric for the current-carrying case is just a
Taub-Kerr-Schild generalization of the previous one.

In the case where there is no current, we can make the following
ansatz for the fields and the metric~\cite{garfinkle,NO}
\begin{equation} \Phi = \varphi (r) e^{in\theta} \ \ \ , \ \ \ \Sigma
= \sigma (r) e^{i\psi} ,\label{vortex} \end{equation}
\begin{equation} ds^2 = -e^{a(r)}dudv+dr^2+e^{b(r)}d\theta^2
\label{metexp} \end{equation}
the only nonvanishing component of $B_\mu$ is taken to be
\begin{equation} B_\theta = B_\theta (r),\label{Btheta}\end{equation}
and, for the time being, we set $A_\mu = 0$ and $\psi =$const. The
equations of motion for the fields then reduce to
\begin{eqnarray} & & B''_\theta + (a' -{b'\over 2})B'_\theta = 4\pi q
\varphi^2 (n+qB_\theta)\nonumber \\
& & \varphi''+(a'+{b'\over 2})\varphi'-e^{-b} \varphi (n+qB_\theta)^2
= {\cal V}_{,\phi} \label{fieldmot}\\
& & \sigma''+(a'+{b'\over 2}) \sigma'={\cal V}_{,\sigma} .\nonumber
\end{eqnarray}
These have to be supplemented with the Einstein's equations for $a(r)$
and $b(r)$. The energy-momentum tensor's nonvanishing components are
\begin{equation} T_{rr} = {1\over 2} \left( \varphi '^2 + \sigma '^2 +
e^{-b} {B'^2_\theta \over 4\pi} \right) - {1\over 2} \left[ e^{-b}
(n+qB_\theta)^2 \varphi^2 + 2 {\cal V}\right]\label{Trr}\end{equation}
\begin{equation} T_{\theta\theta} = {1\over 2}\left[
(n+qB_\theta)^2 \varphi^2 + {B'^2_\theta \over 4\pi} \right] -{1\over
2} e^b \left( \varphi '^2 + \sigma '^2 + 2 {\cal V}\right)
\label{Tthth} \end{equation}
\begin{equation} T_{uv} = g_{uv} {\cal L}(r),\label{Tuv}\end{equation}
where ${\cal L}(r)$ is the Lagrangian
\begin{equation} {\cal L}(r) = -{1\over 2} \left[ \varphi '^2 + \sigma
'^2 + e^{-b} (n+qB_\theta)^2\varphi^2 +e^{-b} {B'^2_\theta \over 4\pi}
\right] - {\cal V} .\label{Lr}\end{equation}

Einstein's equations then take the form~\cite{garfinkle}
\begin{eqnarray} & & 4 R_{uv} e^{-a} = a'' + (a' + {b'\over 2}) a' = 8
\pi G (T_{rr}+e^{-b} T_{\theta\theta})\nonumber \\ & &
-2R_{\theta\theta} e^{-b} = b'' +(a' + {b'\over 2}) b' = 8\pi G
(-4e^{-a} T_{uv} + T_{rr} - e^{-b} T_{\theta\theta}) \label{18}\\ & &
-2 R_{rr} = 2a''+b''+a'^2+{b'^2\over 2} = 8\pi G(-4e^{-a} T_{uv} -
T_{rr} +e^{-b} T_{\theta\theta}).\nonumber\end{eqnarray} Actually, the
third of Eqs.~(\ref{18}) is not independent of the other
two~\cite{garfinkle}. Therefore, Eqs.~(\ref{fieldmot}) and~(\ref{18})
form a system of five coupled differential equations for the five
unknown $B_\theta$, $\varphi$, $\sigma$, $a$ and $b$, with boundary
conditions $\varphi (0) = 0$, $\varphi (\infty) = \eta$, $B'_\theta
(0)=0$, $B_\theta (\infty)=0$, $\sigma '(0) = 0$, $\sigma (\infty)=0$,
$a'(0)=0$, $a(\infty )=0$ and $e^b /r^2 \to 1$ as $r\to 0$.

These equations are far too complicated to solve analytically, even in
flat space ($a=b=0$), where they have been studied only
numerically~\cite{num-witt}. However, it is generally believed that
such solutions should exist, and that, at least for $G\eta ^2 \ll 1$,
their asymptotic properties can be inferred from what is known to be
true in the flat space case~\cite{num-witt}. In this case, the energy
density is concentrated within a region of radius $r_c \sim
\lambda_\phi^{-1/2} \eta^{-1}$ and, for $r\gg r_c$, the energy density
falls off exponentially. Therefore, when gravity is considered, we
expect that the metric should behave as a vacuum solution for $r\gg
r_c$. Static cylindrically symmetric vacuum solutions fall into the
class given by Eq.~(\ref{metricini}), with $c_2 = 0$. If, in addition,
we demand boost invariance along the axis, we must take $m=0$ or
$m=-2$. For $m=-2$, the metric is of Kasner type and has the property
that a circle around the axis at $r\to\infty$ has zero length. Such metric
is considered unphysical from the point of view of strings. For $m=0$,
the exterior metric is just flat space with a deficit angle which is
given by~\cite{garfinkle}
\begin{equation} \Delta \theta = 2\pi (1-\gamma) = 8\pi G\mu + {\cal
O} [(G \mu )^2],\label{18*1}\end{equation}
where
\begin{equation} \mu \equiv 4\pi \int_0^\infty e^{-2a} T_{uv} r dr .
\label{18*2} \end{equation}

The only difference with the Nielsen-Olesen case is that we now have
the condensate $\sigma$ in the core of the string, which will modify
the value of $T_{uv}$ [see Eq.~(\ref{Tuv})], and hence $\mu$, but the
metric is still basically flat space minus a small wedge.

Now, to introduce a lightlike current on the string, we take the
ansatz~(\ref{vortex}) and~(\ref{Btheta}) for $\Phi$, $\Sigma$ and
$B_\mu$, but we introduce a nonvanishing gauge field $A_\mu$ and
$u$-dependent phase $\psi$.
\begin{equation} A_\mu = A(r) p(u) \delta _{u\mu} \ \ \ , \ \ \ \psi =
\psi (u) .\label{Apsi}\end{equation}
As we shall see below, this is consistent with an ansatz for the
metric of the form
\begin{equation} d\widehat s^2 = -e^{a(r)}\left( dudv + H(r,\theta,u)
du^2 \right) +dr^2 + e^{b(r)} d\theta ^2 .\label{20}\end{equation}
The equation for $A_\mu$ following from~(\ref{lagwit}) is
\begin{equation} \nabla _\mu F^\mu _\nu = 4\pi e J_\nu
,\label{21}\end{equation}
where $J_\nu = \sigma^2 (\psi_{,\nu}+eA_\nu )$ is the electromagnetic
current. Eqs.~(\ref{21}) and~(\ref{Apsi}) with the metric~(\ref{20})
require
$$\psi _{,u} = k p(u) $$
where $k$ is a constant, and
\begin{equation} A'' + (a'+{b'\over 2}) A' = 4\pi e \sigma^2
(k+eA).\label{22} \end{equation}
It is interesting to observe that the introduction of the
fields~(\ref{Apsi}) and the new term $g_{uu}$ in the metric do not
affect the form of Eqs.~(\ref{fieldmot}). As we shall see, the
equations for $a$ and $b$ are also unaffected and therefore, the null
current does not change at all the profile of the vortex (unlike the
case of spacelike or timelike currents).

To write down Einstein's equations, we use the fact that, as
mentionned before, the metric~(\ref{20}) is the Taub-Kerr-Schild
generalisation of~(\ref{metexp}). A direct application of Taub's
formalism~\cite{taub} (see, e.g. Eq.(3.23) of Ref.~\cite{gv}) yields
the new Ricci tensor
\begin{equation} \widehat R_{\mu\nu} = R_{\mu\nu} + \widehat R_{uu}
\delta _{\mu u} \delta _{\nu u} ,\label{23}\end{equation}
where $R_{\mu\nu}$ is the old Ricci tensor, given in terms of $a$ and
$b$ by Eqs.~(\ref{18}), and
\begin{equation} \widehat R_{uu} = 2 H R_{vu} +{1\over 2} e^a \Box H
.\label{24} \end{equation}

Also, it is easily verified that the new energy-momentum tensor can be
written as
$$ \widehat T_{\mu\nu} = T_{\mu\nu} +\widehat T_{uu}
\delta _{\mu u} \delta _{\nu u},$$
where $T_{\mu\nu}$ is the old energy-momentum tensor, given by
Eqs.~(\ref{Trr}-\ref{Tuv}), and the new component is given by
\begin{equation} \widehat T_{\mu\nu} = p^2 [\sigma ^2 (k+eA)^2 + {A'^2
\over 2} ] -e^a H {\cal L},\label{25}.\end{equation}

The trace $T^\mu_\mu$ is unaffected, and so the Einstein
equations~(\ref{18}) are unchanged, as promised. However, a new
equation, corresponding to the $uu$ component, has to be considered:
\begin{equation} {1\over 2} e^a [\Box H + H(R_{rr} + e^{-b}
R_{\theta\theta} )] = 8\pi G \widehat T_{uu} .\label{26}\end{equation}
Using~(\ref{18}), we have
$$R_{rr} + e^{-b}R_{\theta\theta} = 32\pi G e^{-a}T_{uv}=-16\pi G
{\cal L}.$$
Hence, from~(\ref{25}) and~(\ref{26}), one arrives at the simple
linear equation for $H$
\begin{equation} {1\over 2} e^a \Box H = 8\pi G p^2 [\sigma ^2
(k+eA)^2 + {A'^2\over 2}].\label{27}\end{equation}
The general solution to this can be obtained as the sum of a
particular solution plus the general solution of the homogeous
equation
\begin{equation} \Box H_h =0.\label{28}\end{equation}
To find the particular solution, we take the ansatz
$$ H = p^2 (u) V(r),$$
which implies
\begin{equation} {1\over 2} e^a \widetilde \Delta V = 8\pi G [\sigma^2
(k+eA)^2 + {A'^2\over 2}],\label{29}\end{equation}
where $\widetilde \Delta = e^{-(a+b/2)}\partial _r (e^{a+b/2} \partial
_r)$ is the Laplacian in the transverse plane. Note that~(\ref{29}) is
just an ordinary differential equation. Note also that outside the
core, $a\to 0$, $\sigma\to 0$ and Eq.~(\ref{27}) reduces to~(\ref{E})
of the previous section. Also, if the thickness of the core is small
with respect to all other relevant length scales, then it is justified
to replace the term $\sigma ^2 (k+eA)^2$ in the right hand side
of~(\ref{27}) by a delta function distribution.

To summarize the results of this section, we can say that the
lightlike current actually decouples from the background fields
forming the static string, since the introduction of the null gauge
field has no effect on $\sigma$, $\varphi$, $a$ and $b$. Its only
effect on the space time geometry can be accounted for by solving the
linear ordinary differential equation~(\ref{29}).

As mentionned before, to the particular solution given by~(\ref{29}),
we can add any solution of the homogeneous equation~(\ref{28}), $H_h
(r,\theta ,u)$. Such solutions can accomodate an additional
gravitational wave, as well as a cosmic string transverse wave
travelling in the same direction as the null current~\cite{gv}.

\section{Geodesics.}\label{geod}

In the stationary case, i.e., when the metric takes the
form~(\ref{metfin}), one can find three constants of motion for
geodesic test particles
\begin{eqnarray} p_- & \equiv & {\dot u \over 2}\nonumber \\
p_+ & \equiv & {\dot v \over 2} + 2 p_- V(r) \label{30} \\
L_\mu & \equiv & (1-4 G\mu) r^2 \dot \theta \nonumber
\end{eqnarray}
where $\mu\equiv {1\over 2} (U'+T') = {1\over 2}(U+T)$ [$\mu$ can be
expressed in terms of underlying fields through~(\ref{18*2})].
In~(\ref{30}), a dot denotes derivative with respect to an affine
parameter $\lambda$. For convenience, we take $\lambda \equiv \tau
/m$, where $\tau$ is the proper time, and $m$ is the mass of the
particle. Using the constraint $\dot x_\mu \dot x^\mu = -m^2$, we have

\begin{equation} \dot r^2 + {L_\mu ^2 \over r^2} + 4 p_-^2 V(r) = -m^2
+ 4 p_+ p_- ,\label{31}\end{equation}
and the problem reduces to the motion of a Newtonian particle in a
potential $V(r)$ given by~(\ref{potential}).

Note, however, that the stationary metric has peculiar asymptotic
properties. In particular, since the potential grows logarithmically
with radius, the motion in the $r$ direction will be bounded even for
light rays ($m^2=0$). Therefore, the problem of asymptotic light
deflection is not well posed because we do not have asymptotic regions
where light rays would propagate along straight lines.

Of course, this extreme behaviour will not arise if, instead of the
stationary metric, we consider a finite pulse of current, because then
the metric is flat before and after the passage of the wave. For
simplicity, we can take a pulse with a step function profile. The
metric is given by~(\ref{pp}) with
\begin{equation} H(u,r) = \Theta (u) \Theta (a-u)
V(r),\label{32}\end{equation}
where $a$ is the duration of the pulse (not to be confused with the
metric function in the previous section). The geodesic equation for
the $v$ coordinate reads
$$ {d\over d\lambda} (4p_- H + \dot v )=4p_- ^2 [\delta (u) - \delta
(a-u)] .$$
Therefore, $p_+$ defined in~(\ref{30}) is only piecewise constant in
this case, undergoing jumps when the particle enters and exits the
wave.

Denoting by $r_1$ the coordinate radius where the wave hits the
particle (at $u=0$) and by $r_2$ the radius at which the particle
emerges from the wave (at $u=a$), we have
\begin{equation} \Delta p_+ \equiv p^{(2)}_+ - p^{(1)}_+ = -p_- \Delta
V, \label{33} \end{equation}
where $\Delta V \equiv V(r_2) - V(r_1)$ and $p^{(1)}_+$ and
$p^{(2)}_+$ are the values of $(\dot v /2)$ before and after the
particle has interacted with the wave. Without loss of generality, we
take $dz/dt=0$ as an initial condition. If $v_z\equiv dz/dt\not= 0$
initially, we can always perform a Lorentz transformation to a new
frame where such component of the velocity will vanish [under this
Lorentz transformation, $H(u,r)\to \kappa H(u,r)$ and $a\to
\kappa^{-1/2} a$, where $\kappa \equiv (1-v_z)/(1+v_z)$]. Then, in the
new frame, $p^{(1)}_+ = p_-$, and, from~(\ref{33}), we find that the
velocity of the particle parallel to the string (after the wave has
past) is
\begin{equation}
\left( {dz\over dt}\right) _2 = {-\Delta V\over 2-\Delta V}
\approx {1\over 2} [V(r_1) - V(r_2)],\label{45b}
\end{equation}
(typically, $\Delta V \ll 1$).
Hence, the particle receives a boost in the $z$ direction whose sign
and magnitude are completely determined by $r_1$ and $r_2$. In
particular, if the particle is initially at rest, it will be attracted
towards the string by the passing wave, so that $r_2 \leq r_1$ [i.e.,
$V(r_2)\leq V(r_1)$] and the boost will be in the positive $z$
direction. On the other hand, a light ray in the $(x,y)$ plane will be
deflected in the $z$ direction by the wave, but the sign of the
deflection depends on whether the interaction with the wave happens
when the ray is approaching the string or when the ray has already
surpassed the string and is moving away from it. In the former case,
the deflection is in the positive $z$ direction, whereas in the latter
it is in the negative $z$ direction.

In the next section, we shall numerically treat the motion of null geodesics
in the spacetime given by~(\ref{32}), but for the rest of this
section, we shall concentrate on a special case which can be treated
analytically, namely the shock wave case~\cite{shock}
\begin{equation} H(u,r) = a \delta (u) V(r).\label{34}\end{equation}
This can be considered as limiting case of~(\ref{32}) when the
thickness of the wave is much smaller than other relevant dimensions
in the problem (such as the impact parameter or the distance of the
wave and of the observer to the string).

It will be convenient to use cartesian coordinates in which the
metric~(\ref{pp}) reads
\begin{equation} ds^2 = -dudv-Hdu^2+dX^2+dY^2
,\label{35}\end{equation}
where $X=r\cos (\gamma \theta)$, $Y=r\sin (\gamma\theta )$, with $-\pi
\leq \theta \leq \pi$, so that there is a wedge of angle $2\pi
(1-\gamma)$ missing in the $(X,Y)$ plane (see fig.~\ref{fig:wedge}).
We shall take
the source to be sufficiently far away so that the incident rays are
parallel to the $X$ axis. The geodesic equations read
\begin{eqnarray} \dot u & = & 2p_- = \hbox{cte}\nonumber\\
& & \nonumber \\
\dot X & = & -{1\over 2} \int {dH\over dX} \dot u du \label{36}\\
& & \nonumber \\
\dot Y & = & -{1\over 2} \int {dH\over dY} \dot u du \nonumber
\end{eqnarray}
and $\dot v$ can be obtained from the constraint
\begin{equation} \dot x_\mu \dot x^\mu = -m^2
.\label{37}\end{equation}

Using~(\ref{34}), we have
\begin{eqnarray} \dot X & = & \dot X_1 -ap_- V'(r_0) {X_0\over r_0}
\Theta (u) \nonumber \\
& & \label{38}\\
\dot Y & = & -a p_- V'(r_0){Y_0\over r_0}\Theta (u)
.\nonumber\end{eqnarray}
Here, $\dot X_1$ is the initial velocity, $V'=dV/dr$, $X_0$ and $Y_0$
are the coordinate where the particle hits the wave, and
$r_0=\sqrt{X_0^2 +Y_0^2}$. For light rays lying in the $(X,Y)$ plane,
we have $\dot X_1 = -2p_-$. From~(\ref{37}),
$$ \dot v = {1\over 2p_-} (\dot X^2 + \dot Y^2),$$
with $\dot X$ and $\dot Y$ given by~(\ref{38}).

Denoting by $\theta$ the deflection in the $(X,Y)$ plane and $\varphi$
the deflection in the $z$ direction, we have, to first order in $V$,
\begin{equation} \varphi \approx {1\over 2} a V'_0 {X_0 \over r_0}
\label{39} \end{equation}
\begin{equation} \theta \approx {1\over 2} a V'_0 {Y_0\over r_0} + \pi
(1-\gamma ) \hbox{sign} (Y_0), \label{40}\end{equation}
where the last term in the r.h.s. of~(\ref{40}) corresponds to the
deficit angle created by the static string.

An interesting question, from the observational point of view, is to
find out what is the effect of the wave on the double image caused by
the static string. As we shall see, the two images undergo apparent
motion, describing (almost) closed trajectories in the sky.

For simplicity, we shall take
\begin{equation} V(r) = \alpha \ln {r\over r_\sigma}, \label{alpha}
\end{equation}
where $\alpha = 8G(U'-T') + 8GI_z'^2 \ln (\bar r/ r_\sigma)$, with
$\bar r$ a cosmological scale. This is a good approximation
to~(\ref{potential}), since in the course of the scattering, $\ln
(r/ r_\sigma)$ changes only by a small percentage.

{}From the geometry of fig.~\ref{fig:wedge}, it is clear that the
deflection angle is given by
$$\theta = {Y_0 + \pi (1-\gamma) X_0 \over \ell + X_0},$$
where $\ell$ is the distance of the observer to the string. Comparing
with~(\ref{40}), we have
\begin{equation} {a\alpha \over 2} {Y_0\over r_0^2} =
{Y_0 -\ell \pi (1-\gamma) \over \ell + X_0}. \label{41}\end{equation}
This equation gives the locus of points at which the rays received by
the observer have been scattered by the wave. From it, we can obtain
$Y_0=Y_0(X_0)$, and, substituting in~(\ref{39}) and~(\ref{40}), $X_0$
can be eliminated to obtain the curve $\varphi = \varphi (\theta )$.
Since~(\ref{41}) is a complicated expression, it is difficult to carry
out this procedure in general. However, we can easily obtain the
answer in two limiting cases.

The first limiting case is when
$$a\alpha \ll 2\ell \pi^2 (1-\gamma )^2.$$
In that case, we can replace $Y_0$ by $\ell \pi (1-\gamma)$
in~(\ref{39}) and~(\ref{40}) to find
\begin{equation} \varphi^2 + \left[ \theta -4G\mu\pi -
{a \alpha \over 16G\mu \ell \pi}\right] ^2 \approx \left[
{a \alpha \over 16G\mu \ell \pi}\right] ^2 ,\label{42}\end{equation}
that is, the double images will describe circles of angular radii
$a \alpha /[ 16G\mu \ell \pi]$.

The second limiting case which can be solved is when the deficit angle
can be neglected, $\gamma = 1$. In that case, we have
\begin{equation}\varphi^2 + \theta ^2 \approx {1\over 2} {a\alpha
\over \ell} . \label{seclim}\end{equation}
The image describes the following trajectory. Initially, it moves
upwards to an angular distance $(a\alpha/2\ell)^{1/2}$ from its initial
position without moving sideways. Then it splits into two images which
describe semicircles of radius $(a\alpha/2\ell)^{1/2}$. Finally, the
two images merge at the bottom of the circle and eventually go back to
the original position. However, for all its beauty, this pattern will
not arise in situations involving realistic cosmic strings. Typically, the
effect of the deficit angle is always comparable, if not much larger,
than the effect due to the current. The reason is that, although a
cosmic string can in principle support an arbitrarily large lightlike
current, it is difficult to think of a mechanism to generate
currents much stronger than the ``critical current'', defined by the
condition $\alpha\sim G\mu$. We shall return to this question in the
concluding section.

Equations (\ref{42}) and (\ref{seclim}) correspond to the case of a
shock wave. If the duration of the wave is
$a\gg$ Max$[G\mu\ell,(a\alpha\ell)^{1/2}]$, then the shock wave
approximation is no longer a good description, since the duration is
large compared with the impact parameter $b$. In such a case, one can
see from Eqs. (\ref{45b}) and (\ref{36}) that to first order in
$\alpha$ the deflection angles are of order $\Delta\theta\sim \alpha$
and $\Delta\varphi\sim \alpha \ln(a/b)$.

\section{Motion of a point source.}

In the previous section we considered the apparent motion of double
images in the sky due to the light deflection caused by an impulsive
wave. However, we neglected the fact that due to the gravitational
field, the observer and the source will undergo geodesic motion, which
will also contribute to the apparent motion. Here, we shall include
this effect, studying the actual motion of a point source as seen from
the observer. Also, we shall consider a wave of finite duration $a$.
We shall assume that initially the source and the observer are both in
a plane orthogonal to the string, and that neither of them is at
infinite distance. We are therefore looking at the particular
configuration shown in Fig.~\ref{fig:not}, where the angles of
observation $\Theta$ and $\phi$ are defined as the usual azimuthal and
polar angles. It turns out that it will be simpler to use, instead of
the forms~(\ref{30}) and~(\ref{31}) for the geodesic equations where
the constant of the motions are explicitely exhibited, the full second
order differential equations directly derived from the
metric~(\ref{pp}) with $H(u,r)$ given by~(\ref{Hur}):
\begin{equation} \ddot r = r\gamma^2 \dot \theta^2 - {1\over 2} {\partial
H\over \partial r} (\dot t - \dot z)^2 ,\label{ddotr}\end{equation}
\begin{equation} \ddot \theta = -{2\dot r\over r} \dot \theta
,\label{ddotth} \end{equation}
\begin{equation} \ddot t = \ddot z = -{1\over 2}{\partial H\over
\partial u}(\dot t - \dot z)^2 - {\partial H\over \partial r}\dot r
(\dot t - \dot z) .\label{ddotz}\end{equation} Here a dot still means
differentiation with respect to the affine parameter $\lambda$ along
the geodesic under consideration and we have explicitely made
use of the conservation of $\dot u$. We consider the case where the
null current is of constant amplitude over a compact support, i.e.,
the function $H(u,r)$ is given by~(\ref{32}).

Now, there is a problem with the set of coordinates we have been
working with to derive Eqs.~(\ref{ddotr}), (\ref{ddotth})
and~(\ref{ddotz}). Indeed, as the wave reaches the observer, the
latter will experience motion in the $z$ direction [see e.g.,
Eq.~(\ref{obsloc}) below]. To avoid this
unphysical ``frame dragging'' effect, it is convenient to pick a
particular coordinate system in which the observer and the source are
at rest. Assuming a free-falling observer of mass $M$
for which we neglect the motion in the radial direction, we have,
according to Eqs.(\ref{pp})
\begin{equation} \dot u \dot v + H \dot u^2 = M^2
,\label{obs}\end{equation}
and since Eqs.~(\ref{30}) tell us that $u$ is proportional to the
proper time $\tau = M \lambda$, we are free to choose $u=\tau$ so that
Eq.~(\ref{obs}) transforms into
\begin{equation} {dv\over du} = 1-H.\label{obsloc}\end{equation}
Incidentally, this equation also tells us that $dz/du = -(1/2)H$
so the motion in the $z$ direction after the wave has past is
\begin{equation}
\Delta z \approx -a V(\ell )/2,\label{incident}
\end{equation}
where $\ell$ is the distance from
the observer to the string. We see that setting
$$\tilde v = v + \int H du \ \ \ \hbox{and} \ \ \ \tilde z = {1\over
2} (\tilde v - u) $$
($u$ being unchanged in the new frame) implies
$$ {d\tilde z \over d\tau} = {d\tilde v\over du} - 1 = \dot v + H -1$$
which vanishes at the observer's and at the source's locations
according to Eq.~(\ref{obsloc}).

Dropping the tildes, we now have the following equation to solve
numerically:
\begin{equation} \ddot t = \ddot z = {1\over 2} {\partial ^2 P \over
\partial r^2} \dot r^2 + {1\over 2} {\partial P \over \partial r}\ddot
r ,\label{newtz}\end{equation}
where
\begin{equation} P \equiv \int H(u,r) du ,\label{P}\end{equation}
while Eqs.~(\ref{ddotr}) and~(\ref{ddotth}) still hold.

That the approximation of neglecting radial motion is valid can be
checked by considering the actual radial displacement experienced by
the observer. According to Eq.~(\ref{ddotr}) with the previous choice
$u=\tau$, and assuming the
observer to be initially at rest, one has
$$\dot r = -{1\over 2} \int ^u {\partial H\over \partial r} \dot u du
\simeq -{1\over 2} {\partial V \over \partial r} \left\{ \matrix{
u\Theta  (u) & \ \hbox{if}\ \ u \leq a \cr a & \ \hbox{for} \ \ u \geq a }
\right. ,$$
where we neglected variations of $V(r)$ with $u$, since this effect is
of higher order. Thus, at most, the correction in $r$ reads
$$r\sim \ell - 2G(U'-T'){a^2\over \ell} - 4GI'^2_z \ln (\ell
/r_\sigma) {a^2\over \ell} \sim \ell ( 1-G\mu {a^2\over \ell ^2} )$$
with $G\mu \sim 10^{-6}$ as discussed earlier. Therefore, since in
practical application, one expects $a\ll \ell$, the motion in the
radial direction can be confidently neglected.

We have investigated numerically the solutions of Eqs.~(\ref{ddotr}),
(\ref{ddotth}) and~(\ref{newtz}) for various values of the parameters
describing the pulse, and the results are displayed on
Figs.~\ref{fig:pht}, \ref{fig:tht} and~\ref{fig:phitheff}. On theses
figures, we have assumed a GUT string of core radius $r_\sigma =
10^{-60}$Mpc, and the internal string parameters have been enhanced to
$4G(U'-T')=4G\mu = 10^{-4}$ and $8G I_z^2 = 10^{-6}$ in order to
magnify the corresponding lensing effect. The string is located at
$\ell = 25$Mpc from the observer and $d=100$Mpc from the source (see
Fig.~\ref{fig:not}), and the duration of the pulse is $a=10$Mpc. This
particular example provides an illustration of a generic situation.

We are mainly interested by what the observer actually sees, that is,
with the notation of Fig.~\ref{fig:not}, we want to know the
variations of the angles $\Theta$ and $\phi$ with time, i.e., we
calculate the values of $\Theta \equiv \arctan (\ell \dot \theta /\dot
r) |_{obs}$ and $\phi \equiv \pi/2 - \arctan (\sqrt{\dot r^2 + \ell ^2
\dot \theta^2}/\dot z) |_{obs}$ as a function of the time of arrival
of each light ray. The predicted observations are the following.
Initially, the observer sees the usual double image (the two parallel
straight lines on Fig.~\ref{fig:tht}) until the wave reaches the first
deflected ray. Then, the two images appear to move away from each
other, while in the same time moving upwards (Fig.~\ref{fig:pht}).
After a time characteristic of the pulse duration (or of the impact
parameter if this is larger than the pulse duration), the images begin
to go down and eventually towards each other, and finally they reach a
new position in the sky as the wave has passed. Put together,
Figs.~\ref{fig:pht} and~\ref{fig:tht} yield Fig.~\ref{fig:phitheff}
which shows the actual apparent trajectory of the source.

The source is seen to follow two open symmetric curves, i.e., the
final location of the source in the sky is not the same as the initial
one. This can be understood as follow: as the wave passes through the
observer and the source, they both reach new positions in space, but
because they were not initially at the same distance to the string
(only case for which we expect closed curves at this level of
approximation), they have not travelled the same distances in the
$z$ direction, the variation in $z$ being $\Delta z\approx -aV(r)/2$
as discussed earlier. Assuming the source to be farther than the
observer implies that the latter has travelled more than the former
which is thus seen below its initial location. More precisely, the
actual angle of observation will be $\phi_0 \sim a\alpha\ln (d/\ell
)/(d+\ell)$, with $\alpha$ as defined previously on Eq.~(\ref{alpha}).
In addition to this effect there is a frame rotation effect: since the
observer moves in both radial and $z$ directions, its frame is being
rotated (parallel transported), so that he will see the source at an
angle $\phi_{final} \sim \phi_0 + a \alpha /\ell$ because of this
rotation. In practice, $\phi_{final}$ will be negligible due to the
large denominators, and the curves will be almost closed.

Another possible way to measure the effect of the wave is through the
redshift of the source: since each light ray experiences a different
gravitational potential, one would expect to observe the source with a
frequency that should vary while the wave passes. For the purpose of
calculating this effect, let us define the redshift ${\cal Z}$ by the
relation
\begin{equation} {\cal Z} \equiv {\nu_{emitted} - \nu_{observed} \over
\nu_{emitted} } = {u_\mu p^\mu \Big| _{emitted} - u_\mu
p^\mu \Big| _{observed} \over u_\mu p^\mu \Big| _{emitted}}.
\label{resdhift} \end{equation}
Here $u_{\mu}$ is the four-velocity of the observer or of the source,
whereas $p^{\mu}$ is the four-momentum of the light ray. We have
$u_\mu p^\mu = \dot t -{1\over 2} \dot r \int du\partial H /\partial r$.
Fig.~\ref{fig:red} shows
this redshift as a function of the time of arrival of the light rays
with, for comparison, the corresponding angles $\Theta$ and $\phi
-\pi/2$, and assuming the same parameters ($d$, $\ell$, $a$, $U$, $T$,
$I$ and $r_\sigma$) as on previous figures. The redshift
(solid line) is seen to be strongly correlated with the deflection
in the $\phi$ direction.
It should be clear that simultaneous measurements of effective motion
of double images (Figs.~\ref{fig:pht} to~\ref{fig:phitheff}) together
with the reshift variability shown on Fig.~\ref{fig:red} would provide
an unambiguous signature of the existence of cosmic strings endowed
with superconducting currents.

The characteristic time scale during which the source is actually
moving in the sky can be evaluated as roughly the wave duration if
this one is large enough (see Fig.~\ref{fig:a}). However, one may expect
astrophysical situations in which $a$ is much smaller than any other
length involved in the problem. In particular, for the idealized shock
wave case of Eq.~(\ref{34}), the observation time should be of the
order of the impact parameter $b\sim G\mu \ell$. In practice, the
actual time during which the motion of the source takes place is thus
$$t_{motion} \sim \ \hbox{Max} \ [a,b],$$ which, in the case of a GUT
string located at a few Mpc from us, would be of a few years. This can
be seen more precisely on Fig.~\ref{fig:a}.$b$ where the time
variability of the angle $\phi$ is shown for various values of the
wave duration $a$: on these curves, it is clear that the
characteristic time in which the signal reaches its maximum is
essentially independent of $a$, and in fact reflects only the
distance of the string to the observer.

Finally, let us ask the question of the dependence of these results
with the various geometric parameters $d$, $\ell$ and $a$.
As long as $d\gg\ell$, the results do not depend strongly on $d$ and
we can take $d\to \infty$ as we did in the previous section.
For fixed $d$ and $a$, the effect decreases with increasing $\ell$.
Also displayed on Fig.~\ref{fig:phitheff} is the
previously discussed result for $\ell = 25, 50$ and 100~Mpc, with a
fixed value $a=10$~Mpc of the wave duration. The effect decreases with
increasing $\ell$, although only by a factor of order 1. Since we are
in the case in which the duration of the wave is much longer than the
impact parameter $b\sim G\mu\ell$, the angular deflection
is of order $\alpha$ in
the $\Theta$ direction and of order $\alpha\ln(a/b)$ in the $\phi$
direction, as discussed at the end of the last section. If the source
is much closer to the string than the observer is, i.e., in the limit
$\ell\gg d$ the angular deflection would be suppressed by a geometrical
factor of order $d/\ell$.

\section*{Conclusions.}

We have derived an exact solution of the Einstein's
equations that describes the gravitational field surrounding
a lightlike current carrying cosmic string.
The current can be of arbitrary shape and time dependence. In the
thin string limit, the exterior metric belongs to the
general class of $pp$-wave solutions. This exterior metric can be
matched with the solution of the Witten model coupled to gravity in
the core of the string. The resulting metric is just a Taub-Kerr-Shild
generalization of the metric of a cosmic string with vanishing
current.

An interesting feature of the solution is that both the
function $A(r)$ characterizing
the electromagnetic field and the function $H(u,r,\theta )$ describing
the gravitational disturbance obey linear equations [see
Eqs.~(\ref{22}) and (\ref{27})]. This fact can be understood
heuristically as a consequence of the Lorentz invariance of the
``seed'' solution and the lightlike nature of the perturbation.
Indeed, as mentioned before, under a Lorentz transformation of
velocity $v_z$ parallel to the string we have $H\to \kappa H$ and
$A\to \kappa^{1/2} A$, where $\kappa = (1-v_z)/(1+v_z)$. Physically, this
reflects the fact that the source is redshifted for $v_z >0$.
Therefore, by making $v_z$ sufficiently close to 1, we can make $H$
and $A$ as small as we wish. Thus, it is not surprising that these
quantities should obey linear equations. This also explains why the
profile of the vortex and the metric functions $a(r)$ and $b(r)$ of
section~\ref{Wit} are unchanged by introducing the lightlike current.

We calculated the motion of light rays
for the case of a pulse current of finite duration $a$ and constant
amplitude characterized by the parameter $\alpha$ given in
Eq.~(\ref{alpha}).
In addition, we have the parameter $G\mu$ characterizing the mass
scale of the cosmic string (typically we have $\alpha\alt G\mu \sim
10^{-6}$ for GUT strings), and the geometric
parameters $d$ and $\ell$, which give the distance from the string
to the source and the observer respectively. We restricted attention
to the case when source and observer are colineal with the string
and are initially in a plane orthogonal to the string axis, obtaining
the following results.

The initial image is the usual double image before the lightlike
current passes, and each one of the double images describes an
(almost) closed loop in the sky as the wave passes. For $d\gg\ell$, the
angular size of the loop is of order $\alpha$ in the direction
perpendicular to the string and of order $\alpha \ln(a/b)$ in the
direction parallel to the string.
Here $b\sim G\mu\ell$ is the typical impact parameter and we are
assuming $a\agt b$.
We have also seen that the images
experience a blueshift and a redshift which is strongly correlated
with the apparent motion in the direction parallel to the string,
and which is also of order $\alpha \ln(a/b)$.
For sources very close to the string, such that
$d\ll\ell$, the magnitude of the angular deflection
is decreased by a geometrical factor roughly of order $d/\ell$,
although for obvious reasons, the redshift effect is unaffected
by this factor.

These results only apply if the duration of the pulse is large
compared with the impact parameter, that is, if
$a\gg $~Max~$[G\mu\ell,(a\alpha \ell)^{1/2}]$. In the opposite limit,
the pulse can be approximated by a delta function shock wave.
Sending the source to infinite distance, the apparent motion
of the double images can be calculated analytically. When the
effect of the deficit angle is dominant with respect to the effect
of the passing wave (more precisely, when $a\alpha\ll\ell(G\mu)^2$),
the double images describe circular motion in the sky, of angular
radius given by $a\alpha/(16\pi G\mu\ell)$.

The time scale needed for the image to go around the curve reflects
the impact parameter as well as the duration of the pulse. The effect
grows to an angular size of order $\alpha$ in a characteristic time
of order $\ell G\mu$. If $a\gg\ell G\mu$ the effect persist for a
time of order $a$, and then decreases again in a time scale of order
$\ell G\mu$. For instance, a GUT string endowed
with a short pulse of GUT lightlike current located at a few Mpc from
us, would
deflect an infinitely far source in a few years time, and along a curve
of angular aperture of the order $10^{-6}$rad, a phenomenon which
should be within the reach of observational detection limits. In fact,
an observation of this kind requires quite particular conditions
due to the scarcity in the number of strings that we expect out to
such distances~\cite{epss}. Nonetheless, we believe that the low
observational probability is largely compensated by the uniqueness of
the signal: such an observation would provide a clear proof of the
existence of cosmic strings, and show that they are of the
superconducting current-carrying kind.

Let us estimate this observation probability more precisely by
proposing a mechanism through which lightlike currents can build in
cosmic string, namely the interaction of such a string with a galaxy,
or any region having an electromagnetic field. In the original Witten
model~\cite{witten}, the currents built in such a way are small
compared with the GUT scale. However, in models with spontaneous
current generation~\cite{scg} one may be able to build up saturation
currents in such a way. If this interaction region is finite, and if
the string can be approximated as straight across it, then a current
will be induced~\cite{witten}, which can be represented by means of a
phase function $\zeta$ along the string as $J^\mu \propto (-\zeta
_{,t},0,0,\zeta _{,z})$. Now, away from the interaction region, where
the external electromagnetic field vanishes, the general solution for
$\zeta$ (to first order in $e$) is~\cite{avv} $$\zeta = \zeta _1 (z-t)
+ \zeta _2 (z+t),$$ and must be constant if no interaction takes
place. So, far from the interaction region, one has two waves of null
current, one travelling in the positive $z$ direction and the other in
the negative $z$ direction, both at the speed of light, and with a
duration of the order of the size of the interaction region. As
mentioned before, the null current does not affect the profile of the
string, and so there is in principle no limit to the amount of
lightlike current that a string can carry.  However, since in the
interaction region the current will be typically spacelike or
timelike, and such currents achieve saturation, it is clear that we
cannot build up lightlike currents whose strength $\alpha$ is much
larger than $G\mu$.

We see therefore that a requirement for observing a string's
lightlike current pulse is that there exist string loops crossing
galaxies say, and whose size would exceed that of the interaction
region. Now we know that the string loop distribution is such
that~\cite{epss}
$$N \sim {\nu \ell^3 \over R t^2},$$
for the number $N$ of loops of size $R$
located in a sphere of radius $\ell$.
Here, $t$ is the age of Universe (10$^{10}$ yr) and $\nu
\approx 0.01$ is a parameter obtained by numerical simulations. The
coherence length of a string due to wiggles is $\Gamma G\mu t$, where
$\Gamma \approx 100$~\cite{Wig} and so we want $R$ to be at least this
value. Moreover, the impact parameter $b\sim G\mu\ell$ being the
typical duration of the effective motion of the source, we require it
to be $b\alt 100$~yr so as to be able to actually measure the effect.
This means that we require $\ell \alt 100$~yr~$/G\mu$. Thus, the
expected number of events can be estimated as $$N\approx {\nu\over
\Gamma} {10^{-24}\over (G\mu )^4} \approx {10^{-28}\over (G\mu )^4},$$
which can be quite small for GUT strings, but otherwise increases
tremendously for lower mass strings, and is of order one for $G\mu\sim
10^{-7}$. This shows that lighter strings would perhaps produce a
detectable (at the level of arc seconds) double image moving according
to our calculations with a correspondingly varying redshift.

It should be mentioned that this work has been involved
with a lightlike current wave separating two regions of the string
where there was no current. Lightlike currents may also exist at the
boundary between spacelike and timelike currents. The time evolution
of such lightlike currents and their gravitational effect is left for
further research. We should also say that
for cosmic strings without currents, time
dependent exact solutions have been studied in the past, e.g. in
Refs.~\cite{gv,ref:19}. In these cases, the positions of the double
images are also time dependent, although the actual apparent motion is
qualitatively different from our case.

Finally, one may consider solutions describing the collision of two
pulses of lightlike current travelling in opposite directions. Before
the collision the metric is a trivial superposition of two solutions
like the one we have considered in this paper, but during and after
the collision the solution will contain interesting non-linear
effects. In particular, since the metric in the impulsive case
resembles very much the Aichelburg-Sexl metric \cite{boostmeth}
(representing the
gravitational field of an ultrarelativistic body), and it is known
that the head on collision of two such shock waves
generates trapped surfaces \cite{peterdeath}, it would be interesting
to study the possibility that the collision of such lightlike currents
would produce black holes.

\vskip1cm

\section*{Acknowledgements}

We are grateful to Peter d'Eath, Allen Everett,
Joao Magueijo, Paul Shellard, Takahiro Tanaka and Tanmay Vachaspati
for suggestions and conversations during the preparation of this
work. This project was supported by SERC grant \#~15091-AOZ-L9.

\nopagebreak

\vskip1cm
\begin{figure}
\caption{Light deflection in the transverse plane in the shock wave
case with the source at infinity.}
\label{fig:wedge}
\end{figure}

\begin{figure}
\caption{The configuration under study as a lightlike-current wave of
duration $a$ passes along a string located at a distance $\ell$ of the
observer and $d$ of the source.}
\label{fig:not}
\end{figure}

\begin{figure}
\caption{Azimutal angle $\phi$ (see Fig.~2) of the
source as seen from the observer as a function of the time when the
light ray reaches the observer. The string parameters [i.e., the
parameters appearing in the potential~(10)] have been set
to $8G I_z^2 = 10^{-6}$, $4G(U'-T') = 4G\mu = 10^{-4}$.}
\label{fig:pht}
\end{figure}

\begin{figure}
\caption{Polar angle $\Theta$ (see Fig.~2) of the source
as seen from the observer as a function of the light ray arrival time
with the same string's parameters as on the previous figure. The
presence of two similar curve simply reflects the double image
expected in the non-carrying case.}
\label{fig:tht}
\end{figure}

\begin{figure}
\caption{Effective observation of the source in the $\varphi - \Theta$
plane (see Fig.~2). Both images in the positive and negative $\Theta$
directions behave symmetrically, and the source is seen as a function
of time to move upwards at first, then down and up again until it
reaches a final location slightly below its initial location.  This
figure has been obtained with a distance from the source to the string
$d=100$~Mpc with a wave duration $a=10$~Mpc, and the distance from the
observer to the string takes the values $\ell =25$~Mpc (full lines),
$\ell =50$~Mpc (dashed lines), and $\ell =100$~Mpc (dotted lines).}
\label{fig:phitheff}
\end{figure}

\begin{figure}
\caption{Redshift of the source (full line) as a function of the time
of arrival of the light rays. The dashed line represents the angle
$\Theta$ and the dotted line the angle $\phi - \pi/2$.}
\label{fig:red}
\end{figure}

\begin{figure}
\caption{Influence of the wave duration $a$ on the resulting
observation. ($a$), the angle $\Theta$ versus time, and ($b$), the
azimuthal angle $\phi$ also versus time; this represents various
values for $a$, namely $a=1$~Mpc (full lines), $a=5$~Mpc (dashed
lines), and $a=10$~Mpc (dotted lines). It can be seen that although
variations in the $\Theta$ direction are strictly correlated with the
wave duration, the same is not true in the $\phi$ direction for which
the rising time is in fact comparable in all cases, reflecting
basically the value of the impact parameter as discussed in the last
section.}
\label{fig:a}
\end{figure}

\end{document}